# Separating Putative Pathogens from Background Contamination with Principal Orthogonal Decomposition: Evidence for Leptospira in the Ugandan Neonatal Septisome

Steven J. Schiff[1,2,3§]*, Julius Kiwanuka[4§], Gina Riggio[3], Lan Nguyen[3,5], Kevin Mu[3], Emily Sproul[3], Joel Bazira[6], Juliet Mwanga[4], Dickson Tumusiime[4], Eunice Nyesigire[4], Nkangi Lwanga[6], Kaleb T. Bogale[1,7], Vivek Kapur[8], James Broach[5], Sarah Morton[1,9], Benjamin C. Warf[10,11], Mary Poss[3,8,§]

[§]Contributed Equally

[1] Center for Neural Engineering, [2]Departments of Neurosurgery, Engineering Science and Mechanics, and Physics, Penn State University, University Park, PA 16802 USA

[3] Departments of Biology and Veterinary and Biomedical Sciences, Penn State University, University Park, PA 16802 USA

[4] Department of Pediatrics, Mbarara University of Science and Technology, Mbarara, Uganda

[5] Dept. of Biochemistry and Molecular Biology, Institute for Personalized Medicine, Penn State University College of Medicine, 500 University Drive, Hershey, PA 17033 USA

[6] Department of Microbiology, Mbarara University of Science and Technology, Mbarara, Uganda

[7] Schreyer's Honors College, Penn State University, University Park, PA 16802.

[8] Department of Veterinary and Biomedical Sciences, Penn State University, University Park, PA 16802 USA

[9] Harvard Neonatal-Perinatal Training Program, Children's Hospital Boston, Boston, MA 02115 USA

[10] Department of Neurosurgery and Department of Global Health and Social Medicine, Harvard Medical School, Boston Children's Hospital, Boston, MA 02115 USA

[11] CURE Children's Hospital of Uganda, Mbale, Uganda

*To whom correspondence should be addressed: Steven J. Schiff, Center for Neural Engineering, W311 Millennium Science Complex, Penn State University, University Park, PA 16802. Tel: 814-863-4210. Email: sschiff@psu.edu

**Abstract**

Neonatal sepsis (NS) is responsible for over a 1 million yearly deaths worldwide. In the developing world NS is often treated without an identified microbial pathogen. Amplicon sequencing of the bacterial 16S rRNA gene can be used to identify organisms that are difficult to detect by routine microbiological methods. However, contaminating bacteria are ubiquitous in both hospital settings and research reagents, and must be accounted for to make effective use of these data. In the present study, we sequenced the bacterial 16S rRNA gene obtained from blood and cerebrospinal fluid (CSF) of 80 neonates presenting with NS to the Mbarara Regional Hospital in Uganda. Assuming that patterns of background contamination would be independent of pathogenic microorganism DNA, we applied a novel quantitative approach using principal orthogonal decomposition to separate background contamination from potential pathogens in sequencing data. We designed our quantitative approach contrasting blood, CSF, and control specimens, and employed a variety of statistical random matrix bootstrap hypotheses to estimate statistical significance. These analyses demonstrate that Leptospira appears present in some infants presenting within 48 hr of birth, indicative of infection in utero, and up to 28 days of age, suggesting environmental exposure. This organism cannot be cultured in routine bacteriological settings, and is enzootic in the cattle that the rural peoples of western Uganda often live in close proximity. Our findings demonstrate that statistical approaches to remove background organisms common in 16S sequence data can reveal putative pathogens in small volume biological samples from newborns. This computational analysis thus reveals an important medical finding that has the potential to alter therapy and prevention efforts in a critically ill population.

# Introduction

Neonatal sepsis (NS) is responsible for over 1 million yearly deaths world-wide [33,13]. In the developing world NS is often treated without an identified microbial pathogen. Pathogen recovery rates in large scale neonatal and infant sepsis in the developing world can be remarkably low (5-10%) using culture techniques [38,6]. In Uganda, two recent reports from high quality referral center laboratories have failed to identify an agent in greater than 60% of NS patients [20,14].

Amplicon sequencing of the 16S ribosomal RNA gene is useful to identify the spectrum of bacteria present in biological and environmental samples. However, even under optimal conditions, contaminating bacteria from recombinant enzymes and reagents are present, and these can dominate the analysis from low-biomass specimens [23]. Numerous approaches have been applied to attempt to correct for contamination in human microbiota samples [11,28]. Nevertheless, at present the identification of dilute putative pathogens within what are normatively sterile body fluids such as blood and cerebrospinal fluid (CSF) remains an open challenge.

In the present study, we applied a novel quantitative approach to separate background contamination from potential pathogens in sequencing data from blood and CSF in a cohort of neonates presenting with clinical evidence of sepsis in Uganda. We present evidence of Leptospira within these infants presenting at 0-28 days following birth. Our findings demonstrate that appropriate statistical modeling to address background contamination from sample handling and library preparation may increase the utility of 16S amplicon sequencing to augment traditional microbiological diagnostic efforts.

# Methods

Ethics Statement

Under Institutional Review Board approval from the Mbarara University of Science and Technology (MUST), Harvard University, and Penn State University (approved protocol #31264EP), the following study was performed. Written consent was obtained from the mothers of neonates meeting clinical criteria for sepsis in both English (the primary national language of Uganda), and Runyankore (the regional language of southwestern Uganda). Further oversight as well as a material transfer agreement was obtained through the Uganda National Council for Science and Technology.

Neonates are a protected population for human studies. In this work, we only examine fluids drawn as a small volume in excess of that required for clinical diagnostics. Although one might consider consenting for blood draws on normal neonates, neither the investigators of this present work nor our institutional ethics boards would be comfortable with such sampling. Furthermore, CSF can never be drawn from normal infants. In the absence of validated immunological tests for the presence of Leptospira infection in neonates, the gold standard remains PCR or DNA sequencing [21]. This study is one in which, although we will rigorously define handling and reagent contamination through appropriate controls, we lack the ability to sample from a control population that is environmentally- and age-matched with our clinically septic neonates. We therefore will create a sophisticated statistical framework to separate handling and reagent contamination from putative pathogens with the following methods.

Clinical Sampling

The Mbarara Regional Referral Hospital is the main teaching hospital for, and is situated adjacent to, the campus of MUST. It is the referral center for southwestern Uganda, and typically admits over 100 cases of presumed NS each year to its pediatric wards.

Eligibility was sought from neonates (less than 1 month of age) whose mothers were at least 18 years of age and who met the following inclusion criteria: 1) Infant with presumed bacterial sepsis with either 1a) fever, lethargy, and poor feeding, or 1b) hypothermia, lethargy, and poor feeding, or 1c) fever, full fontanel, and poor feeding, 2) Infant greater than 2.0 kg weight, and 3) Infant 1 month or less in age. Exclusion criteria were 1) known local infection other than sepsis, 2) known congenital malformation, 3) known cutaneous or gastrointestinal fistula, or 4) known birth trauma such as wounds or fractures. We used these relatively strict clinical criteria to maximize our yield of sick neonates who were likely to have primary microbiological sepsis, as opposed to having signs due to hypoxic-ischemic encephalopathy (HIE) or a known nidus for infection. The period of greatest potential confound for HIE is, of course, in the immediate postpartum period, and such potential confusion decreases progressively for cases presenting after the first few days of neonatal life.

At MUST, procedures on neonates presenting with clinical NS consist of a blood draw for culture and a lumbar puncture. Under no circumstances were procedures performed to retrieve additional volume of blood or CSF for experimental sampling. Withdrawal of blood volumes in the range of 1% for analysis is well below the volumes expected to have any chance of significant effect on the cardiovascular system. Similarly, withdrawal of less than 5% of total CSF volume are routinely withdrawn from neonates without adverse consequences. In order not to expose these infants to any significant risk beyond that of routine medical care, we restricted our study to infants greater than 2 kg. Lower birth weight infants pose technical difficulties with both blood and CSF withdrawal, and have smaller blood and CSF reservoirs to sample. There are, unfortunately, relatively few low birth weight infants who survive in Uganda where the facilities to salvage them are lacking.

Following maternal informed consent, the following samples were collected. Lumbar punctures were performed with sterile disposable styletted neonatal spinal needles using aseptic technique, withdrawing up to 0.6 mL of CSF (less than 5% of CSF volume in a 2 kg infant), allocated for culture and gram stain (0.2 mL), and up to 0.4 mL onto Whatman FTA Indicating Sample Collection Cards (GE healthcare) for genomic analysis. CSF was only withdrawn as free flow from the spinal needle within 1 minute after insertion without suction, and only so long as free flow was obtained.

Blood for all required tests was collected using standard aseptic technique; withdrawing up to 1.0 mL blood (less than 1% of blood volume in a 2 kg infant), of which 0.4 mL was allocated for culture, malaria smear, and HIV testing, followed by up to 0.6 mL for FTA cards.

CSF and blood samples were taken immediately (within 2 hr) upon admission, and prior to antibiotic administration. However, some neonates referred in from a community clinic or health center had received antibiotics prior to referral (31 of 80, 38.7%). No infants were directly admitted from in-hospital delivery settings, and none had an indwelling catheter prior to samples being taken.

In addition, with maternal consent at MUST, a vaginal smear was collected, cultured, and placed on FTA cards. Maternal blood was drawn as well, with consent, for malaria smear, HIV testing (CD4 counts if HIV+), and additional immunological testing. The vaginal bacteriological culture results for the mothers of these NS cases were reported in [14], and the genomic analysis will be performed in the future and reported elsewhere.

These filter paper-based sample collection cards contain cell lysis chemicals, and 100 μL aliquots were placed onto multiple card disks. They were dried overnight in room temperature dessicators, and then sealed in Tyvek pouches with enclosed desiccant pending shipment to the US.

FTA Card Extraction Protocol

Two 6 mm diameter punches were taken from the center of each dried blood spot (and 3 from each dried CSF spot), and placed in a 1.7 mL Eppendorf tube with ATL Buffer and Proteinase K from the Qiagen DNA Micro kit. Punches were taken surrounding the blood or CSF spots (there is an indicator dye on the cards) to serve as negative controls. The card punches were incubated at 56°C for 60 min, vortexing briefly every 10 min. After addition of 300 ul Buffer AL, the tube was transferred to 70C for 10 min. The lysate was transferred to a Qiagen Micro DNA spin column and processed according to protocol. The DNA was eluted in 30 μL 10 mM Tris.

Preparation of libraries for 454 sequencing.

All samples were first screened by PCR for 16S rRNA using universal primers 27F and 907R to determine if there was sufficient sample to generate a library. PCR conditions were as follows: 94°C for 3 min followed by 35 cycles of 94°C for 30 sec to melt the DNA, 60°C for 30 sec to anneal primers, and 72°C for 1 min to synthesize the product. Products that yielded a band of the correct size were advanced for library production.

In order to detect potentially rare pathogenic bacteria in the background of human DNA, we performed 5 replicate PCR each with 1 μL of patient DNA extracted from the filter paper. The first step of the protocol was to produce a template to use for library construction and employed the primers and PCR conditions used above with the exception that 18 cycles of amplification were used. The products were pooled and purified using a Qiagen PCR clean up column.

There are 4 quadrants on the 454 flow cell and contamination can occur between wells. Thus, the second step introduced a novel sequence to the 16S molecule to tag all sequence data as originated in our laboratory. Five reactions were set up using 1 μL of sample from the pooled product of the PCR in step 1. Primers LTR29aF and 700R were used in a PCR reaction with the following conditions: 94°C for 3 min followed by 18 cycles of 94°C for 30 sec to melt the DNA, 58°C for 30 sec to anneal primers, and 72°C for 30 min to synthesize the product. The products of the five PCR were pooled and purified using a Qiagen PCR clean up column.

The third step of the protocol used PCR to add the 454 sequencing adaptors to the 16S fragments. These adaptors incorporate the standard 454 indices but the 3' portion is modified to recognize the unique sequence tag added to the fragment 5' end in step 2. Five reactions were set up using 1 µL of sample from the pooled product of the PCR in step 2. Primers consist of 454 adaptor A and B, and include the universal 16S primer 534R. PCR reaction conditions were: 98°C for 3 min followed by 14 cycles of 98°C for 30 sec to melt the DNA, 57°C for 30 sec to anneal primers, and 72°C for 30 min to synthesize the product and a final extension of 5 min at 72°C. This yields a fragment spanning V1-V4 of the 16S rRNA gene. The amplified fragments were gel isolated and subjected to AmpureBead purification. Each library was quantified by fluorimetry and checked for quality on a BioAnalyzer. Those that passed all quality screens were pooled in equal molar amounts for 454 sequencing. Between 20 and 30 samples were included in each sequencing run.

Taxon identification

The fastQ files from each sequencing run were processed for read quality, the presence of our lab specific sequence tag, and a minimum length of 200 bp. Reads were de-multiplexed and the individual libraries were submitted to the Ribosomal Database Project (Michigan State University) classifier for bacterial identification. The reads classified to the genus level at the 80% confidence level were collated for all patients and blanks.

Amplification of Leptospira rpoB and Streptococcus rnpb

We utilized LeRpoB1F [CCTATGTGGGAACCGGAATGGA] and LeRpoB2R [CGTTTCGTCCTAATGCAAGAGTTC] to amplify a 489bp fragment of the Leptospira RpoB gene. PCR conditions were: 94°C for 3 min followed by 36 cycles of 94°C for 30 sec, 57°C for 30 sec, and 72°C for 30 min. For streptococcus species level identification, we amplified a 330-380 base pair fragment of the Rnase P Beta gene (rnpB) using strF [YGTGCAATTTTTGGATAAT] and strR [TTCTATAAGCCATGTTTTGT ] [32]. PCR conditions were: 94°C for 3 min followed by 36 cycles of 94°C for 30 sec, 56°C for 30 sec, and 72°C for 30 min.

Products were gel isolated. A representative of each was Sanger sequenced. The Streptococcus rnpB product was used for a heteroduplex mobility assay (HMA).

The HMA is a rapid gel based method to identify the sequence similarity between two PCR fragments. If the two fragments are identical they will migrate as a single band after they are melted and allowed to re-anneal. If there are sequence differences between the two fragments then both homoduplexes and heteroduplexes, which represent the reannealed mismatched strands, are formed. Heteroduplexes migrate more slowly on a polyacrylamide gel at distances from the homoduplex roughly proportional to the number of nucleotide differences between the two fragments.

For our assay, we identified a patient who was culture positive for Streptococcus pneumoniae and had detectable Streptococcus 16S rRNA genome sequence, which was most closely related to S. pneumonia in our 454 library screen. Additionally, we chose several patients with detectable 16S rRNA for Streptococcus that was most closely related to S. thermophilus. We amplified the rnpB gene from these patients and sequenced them to confirm species identification. These fragments served as our standards in all assays.

Samples were resuspended in Annealing Buffer (10 mM Tris, 100 mM NaCl, 2mM EDTA) and heated to 94°C for 2 min. The sample is cooled over 10 min to 4°C in a thermocycler and then placed on ice. The samples were resuspended in loading buffer and resolved on a 10% polyacrylamide gel. Standards consisting of the S. thermophilus and S. pneumonia alone mixed together were included on every gel to identify the position of the heterduplex. Gels were visualized by staining with gelRed.

Sequence Controls

We expected environmental contaminants would be present in our samples and took the following steps to identify them. First, we extracted blank cards and prepared libraries from any that amplified a 16S rRNA product. Two such card samples taken from the filter paper surrounding a centralized blood (1) or CSF (1) sample (where the indicator dye was colored) were used as negative controls in the paper. In addition, two negative controls (reaction mix without added template) were included in all PCR reactions. A library was prepared from the only one that yielded a DNA band, which formed our third negative control in our table of read counts (see supplemental information).

Statistical Methods

Fisher's Canonical Discrimination

In 1936 Roland Fisher [8] created a method of multivariable discrimination to help classify data that had more than 1 measurement and that came from more than 1 group of items. Fisher's problem was motivated by related species of flowers. He had petal and sepal length and width measurements of each of 3 species (50 samples each). He was able to find the optimal way of adding these 4 measurement variables together (a linear combination) so that he could clearly show that these sets of measurements could separate and classify each species type. Indeed, the method provides a recipe to measure a new item, weight the measurements, and optimally classify the likely species for such out of sample data [9]. In previous work, we have refined this method, to take into account modern numerical computer algorithms [25 ] (Fisher did all of his work on a hand calculator), and we employ this numerically stable form of discrimination in our analysis of groups of genomic data (blood, CSF, and controls in Fig 1), in this present work. Full details are offered in Mathematical Appendix A.

Modal Reconstruction

Methods of optimal statistical decomposition of sets of data (matrices) into a set of modes has been available for over a century (see review in [26]), and has been applied to data from turbulent fluid mechanics [10] to decisions of court justices [31]. We here employ the technique of singular value decomposition in a novel way. Generally, one retains the most prominent modes in such decompositions to remove smaller noise dominated modes [26]. On the other hand, in the case of contamination dominated analysis of low-mass bacterial microbiomes, one might wish to remove the dominating contaminants that are the most universal feature in the largest modes.

In genomic analysis of microbiomes, contaminants from a wide variety of sources, including the analysis reagents themselves, can dominate the bacterial DNA sequences [23]. We draw on an old theorem original specified in 1907 [27] wherein it was shown how to sum up a set of modes to approximate the underlying original data. We rebuild our data set removing one or more of the largest modes that appear most heavily burdened by contamination. A detailed description of this modal analysis is given in Mathematical Appendix B.

Bootstrap statistics

We examine a set of specific hypotheses in our statistical analysis of such modal data. We can randomize by patient – swapping the labeling of data by permuting the codes of the patient samples. We can randomize by genomic taxa – permuting the identification of the taxonomic matches. Lastly, we can assume that our entire data set is random noise, and permute the entire data matrix (see Supplementary S1 Table) where all points in the matrix are exchanged with points randomly chosen from any patient or taxanomic designation. We apply such bootstrap statistics in the analysis of our data.

**Results**

To develop a comprehensive understanding of the bacterial composition of the neonatal septisome in Ugandan infants, we sequenced a fragment of the bacterial 16S rRNA gene using Roche 454 technology from samples of blood and CSF stored on filter paper cards. Of the samples from 80 infants, 65 blood and 27 CSF samples had polymerase chain reaction (PCR) detectable 16S DNA. Five of the 80 infants (6%) were born to HIV+ mothers (similar to the general population rate in this region).

To control for contaminants introduced from sample handling and recombinant reagents, we included specimens from filter paper cut from around (the periphery) of blood and CSF specimens. All PCR amplification steps included a reagent control to which no patient DNA was added, and the solitary reagent control that yielded a 16S band after PCR was sequenced. The results of these 3 handling and reagent contamination controls are given in supplementary information (S1 Table). (see results of control sequencing in supplementary S1 Table).

We first assessed the data for potential contaminating organisms. The patterns of bacteria associated with NS and those from background contamination should have different distributions within the data. In Fig 1A, we show the 131 organisms that could be assigned with 80% confidence at the genus level using the Ribosomal Database classifier [34] and their presence in each patient sample. These data demonstrate that some organisms were ubiquitous among the patient samples and hence were putative contaminants.

***

**Fig 1. The characterization of the dataset and modes. A:** The graphical representation of read counts sorted by columns of total reads from left to right in descending order for 131 genus identifications in 95 samples. Color map is scaled to amplify the lowest 1% of read counts. **B:** Fisher's canonical linear discrimination demonstrates the optimal linear combinations of the read counts ($Z_1$ and $Z_2$) that separate samples from blood, CSF, and controls. Two of the 3 control samples overlap in the plot. Group means are large symbols. **C:** First 10 eigenmodes from principal orthogonal decomposition and total energy (E fraction) accounted for by summing modes progressively from left to right. Only the first 10 columns are plotted in each mode. **D:** The weighting of each mode (eigenvalues) are shown, as well as the tolerance for insignificance (dashed line) below which eigenvalues are not resolvable. **E:** Composition of the modes in terms of their representative genera sorted in descending order as blue, green, and red.

***

We first asked whether there was any signal in our data that could distinguish samples and controls. Fisher's canonical linear discrimination tests whether there are correlations between versus within putative groups of variables that can discriminate groups based on the correlation structure of the variables [8]. We applied Fisher's canonical linear discrimination to the read counts from all of the samples and controls and find that the pattern of bacterial distribution among blood, CSF, and control samples are readily discriminable (Wilks' lambda chi square $p<0.007$, plug-in error rate 0.01, Fig 1B, see reference [9]). These statistics demonstrate that there is signal to discriminate patient samples and controls, indicating that it is extremely unlikely that all 16S reads were the result of random contamination.

We assume that the correlation patterns among bacteria contributing to background contamination are independent of patterns of invasive pathogenic species causing disease. The matrix of bacterial genera in each patient sample can be decomposed into a weighted set of orthogonal patterns using principal orthogonal decomposition, which has shown broad utility in fields as diverse as fluid dynamics [10], legal decisions [31], and neurophysiology [24]. In this approach, we ask what is the most statistically significant projection of all of the data, generating a pattern or mode that is composed of linear combinations of the taxonomic assignments represented on the abscissa in Fig 1A. We then produce a second mode that is the next best projection, etc. Such modes generate a weighting (an 'eigenvalue') that we can use to gauge the percent of a mode within an entire dataset as an energy.

In the combined blood and CSF specimens, 99% of the energy of the data signal is accounted for by 3 patterns (modes) of the data (Fig 1C), the largest of which is dominated by Ralstonia, a common contaminant and rare opportunistic pathogen [12] (Figs 1D, 1E blue). The second mode is comprised of Streptococcus sp, Corynebacteria sp, and E. coli (Fig 1E, green). Leptospira species dominate the third largest mode (Fig 1E, red).

We anticipate that the interactions of putative bacterial contaminants, or the interactions of pathogens in polymicrobial infections, will demonstrate correlations within these patterns. Random matrix theory was developed initially to help explain the interactions between elements of complex nuclei [37], and similarly have been used to study the interactions of stocks that increase and decrease value together in financial analysis [22]. We employ a random matrix approach to quantify the mode significance, randomizing the data matrix (Fig 1A) by patient (rows), genera (columns), or full randomization permuting all read counts among patients and genera (Fig 2B), generating a variety of null hypotheses. Bootstrap ensembles of 1000 separate randomizations from the original data demonstrate that for all samples (Fig 2A), the first mode, dominated by Ralstonia, is the only highly significant mode when compared with full or bacterial (not shown) randomization.

***

**Fig 2. Hypothesis testing for modes using random matrices. A:** Random matrix bootstrap ensemble distribution for all samples showing the mean (black solid line) and ± 1 standard deviation (blue dotted line) for 1000 randomizations of all matrix values, and original data set eigenvalues (red asterisks). **B:** Graphical representation of a randomization of Fig 1A. **C:** All samples with mode 1 removed, and comparable mode composition in **D**. **E** shows eigenvalue distribution for blood samples only, with **F** mode composition. **G** shows eigenvalues for blood with mode 1 removed, and **H** mode composition. **I** illustrates the probabilities of the first 3 modal eigenvalues from **G** and **H** illustrating the significance of dominant Leptospira mode from **H** (similar results randomizing only by bacterial type not shown).

***

In typical uses, one might filter noise contamination in data by removing all small modes below a certain size [19]. But in our case, we wish to do the opposite – removing large modes that represent putative contaminants and then rebuilding the data set by summing the remaining modes to evaluate potential pathogenic bacteria. The mathematics to approximate a matrix with a subset of modes in this way was described by Schmidt in 1907 [27], and we employ that approximation to reconstruct the data set without the first mode (a detailed discussion of such modal sums can be found in [26]). After removing the Ralstonia-dominated mode from all samples, there are two significant modes (Fig 2C): one dominated by Streptococcus sp. and one dominated by Leptospira sp (note the two red asterisks above the randomized confidence limits in Fig 2D). We confirmed that 68 of 74 Streptococcal assignments were S thermophiles (a common contaminant) using a heteroduplex mobility assay (HMA, see Methods). We now need to address whether Leptospira is a putative pathogen in these neonates.

Evidence for Neonatal Leptospira

We first assessed whether the distribution of Leptospira was random in the blood versus the CSF of patients, which would be expected if it were a contaminant. Of the 40% (32/80) of samples with identified Leptospira 16S rRNA, 31 patients had evidence of Leptospira only in the blood. One patient had Leptospira present in both blood and CSF, and one in CSF only. A chi-square analysis of specimens with and without evidence of Leptospira in blood vs CSF rejects the null hypothesis of random contamination (chi-sqare = 32.1, df = 1, p<0.001). Importantly, unlike the case with Ralstonia and Streptococci (S1 Table), Leptospira was not detected in any of our control cards analyzed from blank regions of patient and laboratory cards (n=3), or negative control PCRs. Since only 2 patients had Leptospira in CSF, we examine the modal patterns of only blood. We find that blood has a single significant dominant mode (Fig 2E), which is Ralstonia (Fig 2F). If we remove this first mode from blood, the single remaining dominant mode (Fig 2G) is a nearly pure Leptospira mode (Fig 2H). Examining the distribution of the magnitudes of these modes (eigenvalues) from 1000 bootstrap permutations, this solitary Leptospira mode is significant at $p< 0.001$ (Fig 2I).

Our data were only classified at the 80% confidence level to the genus level. Because 454 read length is variable (< 500 bp), we extracted the longest sequences of those classified as Leptospira from each patient and submitted them to basic local alignment search (BLAST) [1] to ascertain the closest match. Thirty-one (31) NS patient sequences matched equally well to either of the closely related pathogenic species L. broomii or L. inadai [15]. To provide additional support for species designation, we amplified the RpoB gene using PCR from blood samples 19, 21, and 27, and submitted it for Sanger sequencing. The data confirm the placement of Leptospira into the broomi/inadai cluster.

Of the 32 patients with evidence of Leptospira 16S DNA, one had a positive CSF culture for S. pneumonia. Streptococcal 16S DNA was present in both CSF and blood of this patient and we confirmed that the organism was S. pneumonia by sequencing the RnpB gene. Eight patients were culture positive for Staph aureus, which was not identified in our sequence data.

The timing of the presentation of the NS patients with evidence of Leptospira revealed 4/32 (12.5%) presenting with evidence of sepsis on the day of birth (day-0). These day-0 sepsis patients imply in utero infection, and all had high read counts for Leptospira (704-3951). There were 19/32 (59%) patients presenting during the first week, days 1-6 suggesting peripartum infection, and 9/32 (28%) late infections, presenting with NS on days 9-29 after birth suggesting environmental sources.

Evidence for other organisms

Combining culture and sequencing results supports the possibility that there might be polymicrobial underpinnings of NS in this setting [17]. Of the 32 patients with sequence evidence for Leptospira, 8 (25%) had positive blood cultures for S. aureus (7); four of these patients had low read counts of Staphylococcal taxa in our 16S data. One patient with sequence evidence of Leptospira in the blood was positive by culture for S. pneumonia in the CSF. We detected S. pneumoniae in the 16S data and confirmed it by PCR to the Streptococcal ribosomal polymerase gene (rPoB). Three (3) patients with Leptospira sequences had evidence based on 16S gene detection for Acinetobacter, which can be a virulent nosocomial pathogen, as well as a frequent hospital and reagent contaminant. Our prior work on postinfectious hydrocephalus [36] presenting in survivors of NS provided evidence of Acinetobacter species infection in Ugandan neonates [17].

There were 10/160 samples that yielded coagulase-negative Staphylococcal sequences at the genus level, and our assumption is that in the absence of indwelling catheters or known immunocompromise, that the most likely explanation of such a distribution of coagulase-negative Staphylococcal genera is due to recovery of commensal skin organisms.

**Discussion**

Mortality rates of 34/1000 births [33] from NS in sub-Saharan Africa have been difficult to control, in part because of the low rate of pathogen identification. In addition, the long-term sequelae in the survivors of NS, such as postinfectious hydrocephalus [36,17], may add an effective 10% mortality to these figures (hydrocephalic mortality occurs after the neonatal period) [35].

Leptospirosis is presumed to be the most common zoonotic disease in the world [26]. It is present in East African communities at high rates. A recent study in Tanzania demonstrated a seroprevalence of 15.5%, with higher rates for people with extensive contact with cattle [29]. Leptospira is enzootic in cattle and buffalo herds in Western Uganda [2], geographically coincident to where our patients live, in addition to dogs [18], goats and hippopotami [3]. Although infants do not have contact with buffalo or hippopotami, they often live in intimate contact with domestic cattle, goats and dogs [17].

Our data demonstrate multiple instances of Leptospira in blood samples of infants with NS at birth. Leptospira crosses all tissue barriers, including the placenta, and maternal infection during pregnancy has been typically associated with miscarriage and stillbirth; nevertheless, there have been rare documented cases of congenital cases of Leptospirosis with survival following treatment [30]. Our data reveals evidence of neonatal Leptospira consistent with congenital vertical transmission, peripartum infection during the first week of life, and later environmental infections during weeks 2-4 of life. Such a distribution of case presentations speaks to the ubiquity of this organism in both animal and human hosts in this setting.

Leptospira species are broadly susceptible to the antibiotics typically used when neonates present with NS [7,4]. Nevertheless, the extreme difficulty in identifying this organism using bacteriological culture can lead to a lack of adequate antibiotic coverage even in extremely well resourced settings [39].

Of the 26 culture positive patients, 17/26 (65%) had sequence evidence of a pathogen, but only 5/26 (19%) had sequence evidence congruent with the culture organism type. If we exclude S. aureus, for which we had no sequence confirmation, there were 11 culture positive patients remaining, of which there was sequence congruence in 5/11 (45%). Of the 54/80 (67.5%) culture negative patients, there was sequence evidence of a pathogen in 32/54 (59%). These included sequence identification of Acinetobacter baumanii (6), and sequence, PCR, and HMA confirmation S. agalactiae (1) and S. pneumoniae (2). Thus the addition of bacterial sequence to bacterial culture data suggests evidence that increases the potential diagnostic yield from our prior bacteriological analysis from 26/80 (32.5%) to at least 49/80 (61%), an increase of 23/26 (88%) across our patient population.

Despite recent encouraging results demonstrating that individual actionable diagnostic information might come from DNA sequencing [39], our findings do not achieve such promise at the individual patient level. Our results highlight the integral role that rigorous analytic approaches to 16S or other sequencing methods may have in identifying organisms that do not grow in routine culture conditions, such as Leptospira, and in confirming the identity of those organisms that are identified by typical bacteriological methods. Our sequencing further appears useful to help differentiate genus and species of organisms for which comprehensive bacterial biochemical testing is not available, to provide an informed estimation of bacterial spectrum in settings when the lack of a priori knowledge about the relevant pathogen spectrum would otherwise render test panel selection (such as PCR) incomplete, and to raise questions regarding potential false positives if genetic information is unable to confirm culture and biochemical identification.

A combined diagnostic approach consisting of organism culture and computational metagenomics may substantially improve our characterization of the neonatal septisome. As part of this methodology, rigorous statistical analyses of data, such as what we employ here, are needed to address the significant problem of bacterial contamination that occurs at all steps of sample collection and processing. Furthermore, although we have confined our present analysis to potential bacterial causes of neonatal sepsis, future strategies will need to embrace potential non-bacterial causes of sepsis. Only once we have more comprehensively defined the spectrum of the underlying microbial etiologies of these infections, can we more effectively undertake the task of addressing the routes of infection to better prevent NS in settings where it remains out of control.

**Acknowledgements**

This work was supported the generosity of the endowment funds of Harvey F. Brush, the Rose and Sydney P. Schiff Fund for Neural Engineering, a grant from the Penn State Clinical and Translational Sciences Institutes, and US NIH grant 1DP1HD086071.

**Figure Legends**

**Fig 1. The characterization of the dataset and modes. A:** The graphical representation of read counts sorted by columns of total reads from left to right in descending order for 131 genus identifications in 95 samples. Color map is scaled to amplify the lowest 1% of read counts. **B:** Fisher's canonical linear discrimination demonstrates the optimal linear combinations of the read counts (Z1 and Z2) that separate samples from blood, CSF, and controls. Two of the 3 control samples overlap in the plot. Group means are large symbols. **C:** First 10 eigenmodes from principal orthogonal decomposition and total energy (E fraction) accounted for by summing modes progressively from left to right. Only the first 10 columns are plotted in each mode. **D:** The weighting of each mode (eigenvalues) are shown, as well as the tolerance for insignificance (dashed line) below which eigenvalues are not resolvable. **E:** Composition of the modes in terms of their representative genera sorted in descending order as blue, green, and red.

**Fig 2. Hypothesis testing for modes using random matrices. A:** Random matrix bootstrap ensemble distribution for all samples showing the mean (black solid line) and ± 1 standard deviation (blue dotted line) for 1000 randomizations of all matrix values, and original data set eigenvalues (red asterisks). **B:** Graphical representation of a randomization of Fig 1A. **C:** All samples with mode 1 removed, and comparable mode composition in **D**. **E** shows eigenvalue distribution for blood samples only, with **F** mode composition. **G** shows eigenvalues for blood with mode 1 removed, and **H** mode composition. **I** illustrates the probabilities of the first 3 modal eigenvalues from **G** and **H** illustrating the significance of dominant Leptospira mode from **H** (similar results randomizing only by bacterial type not shown).

Supporting Information

S1 Table Title: Full dataset on which paper is based.

S1 Table Caption:

Table of number of reads matching taxanomic identification at the genus level or better. Group 1 represents blood, group 2 CSF, and group 3 control samples. Columns are sorted with the total number of reads to a genus for all samples are largest on the left, and decreasing progressively to the right.

**Mathematical Appendix**

A. Fisher's Canonical Discrimination

Discrimination was performed using methods described in detail in [25]. In brief, the data matrix was assembled into a matrix $\mathbf{Y}$, where the rows are patient or control samples (95), and the columns represent the genus resolved organisms (131). We partition the matrix in blocks of rows corresponding to blood (group 1), CSF (group 2), and controls (group 3), yielding upper, middle, and lower matrices $\mathbf{Y}_1$, $\mathbf{Y}_2$, and $\mathbf{Y}_3$ (see Supplementary S1 Table). The multivariate means of these matrices were computed as $\bar{\mathbf{y}}_j = \frac{1}{N_j}\sum_{i=1}^{N_j} \mathbf{y}_{ji}$, where $\mathbf{y}_{ji}$ are $i$ rows (samples) from the matrix $\mathbf{Y}_i$ for groups $j=1,2,3$. The corresponding covariance matrices are

$$\Psi_j = \frac{1}{N_j}\sum_{i=1}^{N_j}(\mathbf{y}_{ji}-\bar{\mathbf{y}}_j)^T(\mathbf{y}_{ji}-\bar{\mathbf{y}}_j),$$

where T indicates transpose, and the full covariance matrix for the entire data set is $\Psi_{total} = \frac{N-1}{N^2}\sum_{i=1}^{N}(\mathbf{y}_i-\bar{\mathbf{Y}})^T(\mathbf{y}_i-\bar{\mathbf{Y}})$. Pooled covariance within groups, $\Psi_{within}$, was calculated as

$$\Psi_{within} = \frac{1}{N_1+N_2+N_3}\left[(N_1-1)\cdot\Psi_1 + (N_2-1)\cdot\Psi_2 + (N_3-1)\cdot\Psi_3\right]$$

and the between group variance is thus

$$\Psi_{between} = \Psi_{total} - \Psi_{within}$$

Fisher [8] recognized that for any linear combination $\mathbf{z} = \mathbf{Yb}$, where $\mathbf{b}$ is a column vector of coefficients, that the variance, $\text{var}[\mathbf{z}]$, is

$$\text{var}[\mathbf{z}] = \mathbf{b}^T\Psi_{total}\mathbf{b} = \mathbf{b}^T\Psi_{within}\mathbf{b} + \mathbf{b}^T\Psi_{between}\mathbf{b},$$

and that separate groups $j$ implies that $\Psi_{total} \gg \Psi_{within}$.

Our goal is to find the discrimination function $Z(\gamma)$ that best emphasizes the between with respect to the within covariances, or in other words to maximize the ratio

$$\frac{\mathbf{b}^T\Psi_{total}\mathbf{b}}{\mathbf{b}^T\Psi_{within}\mathbf{b}} = 1 + \frac{\mathbf{b}^T\Psi_{between}\mathbf{b}}{\mathbf{b}^T\Psi_{within}\mathbf{b}} = 1+\alpha$$

over all vectors of coefficients $\mathbf{b}$. Then $Z(\gamma) = \mathbf{Yb}$ will be the optimal discriminator,

and the maximum $\alpha$ will quantify the excess between covariance, $\Psi_{between}$.

Fisher's insight [9] was that this maximization can be achieved with a simultaneous spectral decomposition of $\dfrac{\mathbf{b}^T \Psi_{between} \mathbf{b}}{\mathbf{b}^T \Psi_{within} \mathbf{b}}$

$$\max\left[\frac{\mathbf{b}^T \Psi_{between} \mathbf{b}}{\mathbf{b}^T \Psi_{within} \mathbf{b}}\right] \Rightarrow \frac{\mathbf{b}^T \mathbf{H}\Lambda\mathbf{H}^T \mathbf{b}}{\mathbf{b}^T \mathbf{H}\mathbf{H}^T \mathbf{b}} = \frac{\mathbf{b}^T \Lambda \mathbf{b}}{\mathbf{b}^T \mathbf{b}} = \alpha$$

Maximizing $\alpha$ leads to $k=1,\ldots,m$ orthogonal linear combinations $\mathbf{z}_k = \mathbf{Y}\gamma_k$, where $\gamma_k$ are the columns of $(\mathbf{H}^T)^{-1}$. $\Lambda$ is a diagonal matrix, whose values are $\lambda_1 \geq \ldots \geq \lambda_m > 0 = \lambda_{m+1} = \ldots = \lambda_p$, where $p$ are the number of variables, in our case 131. Thus there are $m$ canonical discrimination functions, $\mathbf{z}_k$ which are linear combinations $\mathbf{Y}\gamma_k$ corresponding to the non-zero eigenvalues $\lambda_{1,\ldots,m}$.

In [25], we used singular value decomposition (SVD)

$$\mathbf{Y} = \mathbf{U}\mathbf{S}\mathbf{V}^T$$

to finding the optimal discrimination functions. We change coordinates to simplify the discrimination problem. Let $\Psi_{within} = \mathbf{U}\mathbf{S}\mathbf{U}^T$ be the SVD of $\Psi_{within}$, where $\mathbf{S}$ is diagonal, and $\mathbf{U}$ appears twice because covariance matrices are symmetrical. Define a new variable $\mathbf{v} = \mathbf{U}\mathbf{S}^{1/2}\mathbf{U}^T\mathbf{b}$, or equivalently $\mathbf{b} = \mathbf{U}\mathbf{S}^{-1/2}\mathbf{U}^T\mathbf{v}$. In terms of $\mathbf{v}$,

$$\alpha = \frac{\mathbf{v}^T \mathbf{U}\mathbf{S}^{-1/2}\mathbf{U}^T \Psi_{between} \mathbf{U}\mathbf{S}^{-1/2}\mathbf{U}^T \mathbf{v}}{\mathbf{v}^T \mathbf{U}\mathbf{S}^{-1/2}\mathbf{U}^T \Psi_{within} \mathbf{U}\mathbf{S}^{-1/2}\mathbf{U}^T \mathbf{v}} = \frac{\mathbf{v}^T \mathbf{U}\mathbf{S}^{-1/2}\mathbf{U}^T \Psi_{between} \mathbf{U}\mathbf{S}^{-1/2}\mathbf{U}^T \mathbf{v}}{\mathbf{v}^T \mathbf{v}}.$$

This is a much better coordinate system in which to do the maximization [25]. Since the length of $\mathbf{v}$ scales out of the ratio, it is equivalent to maximize over unit vectors $\mathbf{v}$. We know that in general, the maximum of $\mathbf{v}^T \mathbf{A}\mathbf{v}$ for a symmetric matrix $\mathbf{A}$ is reached for $\mathbf{v} = \mathbf{v}_1$, the first singular vector of $\mathbf{A}$. Furthermore, the maximum subject to being orthogonal to $\mathbf{v}_1$ is $\mathbf{v}_2$, the second singular vector of $\mathbf{A}$, etc. So the maximization is solved by taking the SVD

$$\mathbf{U}\mathbf{S}^{-1/2}\mathbf{U}^T \Psi_{between} \mathbf{U}\mathbf{S}^{-1/2}\mathbf{U}^T = \mathbf{V}\mathbf{A}\mathbf{V}^T$$

and the maximum $\alpha$ is $\mathbf{v}_1^T \mathbf{V}\mathbf{A}\mathbf{V}^T \mathbf{v}_1 = \lambda_1$, the largest singular value from $\mathbf{A}$. Converting back to $\mathbf{b}$-coordinates, the optimal $\mathbf{b}$, called the *first canonical variate*, is

$$\mathbf{b}_1 = \mathbf{US}^{-1/2}\mathbf{U}^T\mathbf{v}_1$$

which is the first column of $\mathbf{US}^{-1/2}\mathbf{U}^T\mathbf{V}$. The second column $\mathbf{b}_2$ of $\mathbf{US}^{-1/2}\mathbf{U}^T\mathbf{V}$ is the second canonical variate, and so on. The $m$ canonical variates $\mathbf{b}_1,\ldots, \mathbf{b}_m$, are the $m$ columns of $\mathbf{US}^{-1/2}\mathbf{U}^T\mathbf{V}$. They provide the coefficients of $m$ *canonical discrimination functions* $\mathbf{Z}_i(\gamma) = \mathbf{Y}\mathbf{b}_i^T$. We plot the first 2 columns of $\mathbf{Z}_i$ in Fig 1A.

For each multivariate data vector $\mathbf{Y}$, the transformed vectors $\mathbf{z}$ have means $\mathbf{u}$ and normal p-variate distributions $f(\mathbf{z})$. Prior probabilities $\pi_j$ are determined from the fraction of total samples within group $j$, $\pi_j = N_j/N$. The posterior probability $\pi_{jz}$ is the probability that for a given value of $\mathbf{z}$, that the data came from group $j$ of $n$ groups

$$\pi_{jz} = \frac{\pi_j f_j(\mathbf{z})}{\sum_{k=1}^{n} \pi_k f_k(\mathbf{z})}, \quad k = 1,\ldots,n$$

A suitable approximation to $\pi_j f_j(z)$ is given by $\exp[q(z)]$ where $q(\mathbf{z}) = \mathbf{u}_j^T\mathbf{z} - \frac{1}{2}\mathbf{u}_j^T\mathbf{u}_j + \ln \pi_j$ [9]. The highest posterior probability among all possible groups is the predicted group membership used in our calculation of plug-in error rate reported in the text referring to Fig 1A.

A normal theory method to test for the significance of discrimination is to examine the magnitude of the eigenvalues of $\Lambda$ above. We make use of Wilks' statistic, $W$. After calculating the log likelihood ratio as $\text{LLRS} = N\sum_{i=1}^{m} \ln(1+\lambda_i)$, where $\lambda_i$ are the diagonal entries of $\Lambda$, $W = \exp\left[-\frac{1}{N}\text{LLRS}\right]$. A poor discrimination yields small eigenvalues $\lambda$, and $W$ approaches 1. Good discrimination yields large eigenvalues, and $W$ becomes small. Since $W$ is chi-squared distributed, we can calculate confidence limits that the discrimination shown in Fig 1A is significant as described in the text.

<u>B</u>. Modal Reconstruction

For an arbitrary matrix Y, the SVD is

$$\mathbf{Y} = \mathbf{U}\Lambda\mathbf{V}^T$$

where $\mathbf{U}$ is matrix of orthogonal columns of sample eigenmodes, $\Lambda$ a diagonal matrix of eigenvalues, $\lambda_i$, and $\mathbf{V}$ a matrix of orthogonal columns of genus modes.

We make use of the fact that the sum of the outer products of the columns of $u_i$ and $v_i$, weighted by their eigenvalues $\lambda_i$, are equal to the original data matrix $\mathbf{Y}$

$$\mathbf{Y} = \sum_{i=1}^{n} u_i \lambda_i v_i^T$$

where n are the total number of modes, in this case, 131.

We employ a definition of tolerance for eigenvalue size in Fig 1D which is the product of the largest singular value, $\lambda_{max}$, times the machine precision of the computer [26]. Eigenvalues smaller than the tolerance are considered computationally meaningless.

Using the Schmidt approximation theorem [27], one typically reconstructs a data matrix **Y** with a subset of modes as

$$\mathbf{Y} = \sum_{i=1}^{m} u_i \lambda_i v_i^T$$

where i=1…m represent the m largest eigenvalues.

.

Such approximations are generally done by retaining the largest modes because a small subset may contain a disproportionate amount of the variance or energy in the signal, $E$, defined as

$$E = \sum_{i=1}^{m} \lambda_i \bigg/ \sum_{i=1}^{n} \lambda_i$$

where m<n.

In typical use, one anticipates that the largest *m* out of *n* components contains the signal of interest. In our case, our signal contains a mixture of background contamination and potential pathogens, and the background may dominate the eigenspectrum. We therefore reconstruct our data set without inclusion of putative background modes, as

$$\mathbf{Y} = \sum_{i=1}^{n-k} u_i \lambda_i v_i^T \ .$$

where we here remove *k* of the largest modes. We employ this formulation in Fig 2 when we remove background modes. A much more detailed description of the reconstruction of such data sets from sums of modes can be found in chapter 7 of Schiff 2012 [26].

Our reconstruction is focused on modes with large eigenvalues. Recent work exploring smaller eigenvalues in undersampled complex biological data can be found in [5].

Figure 1

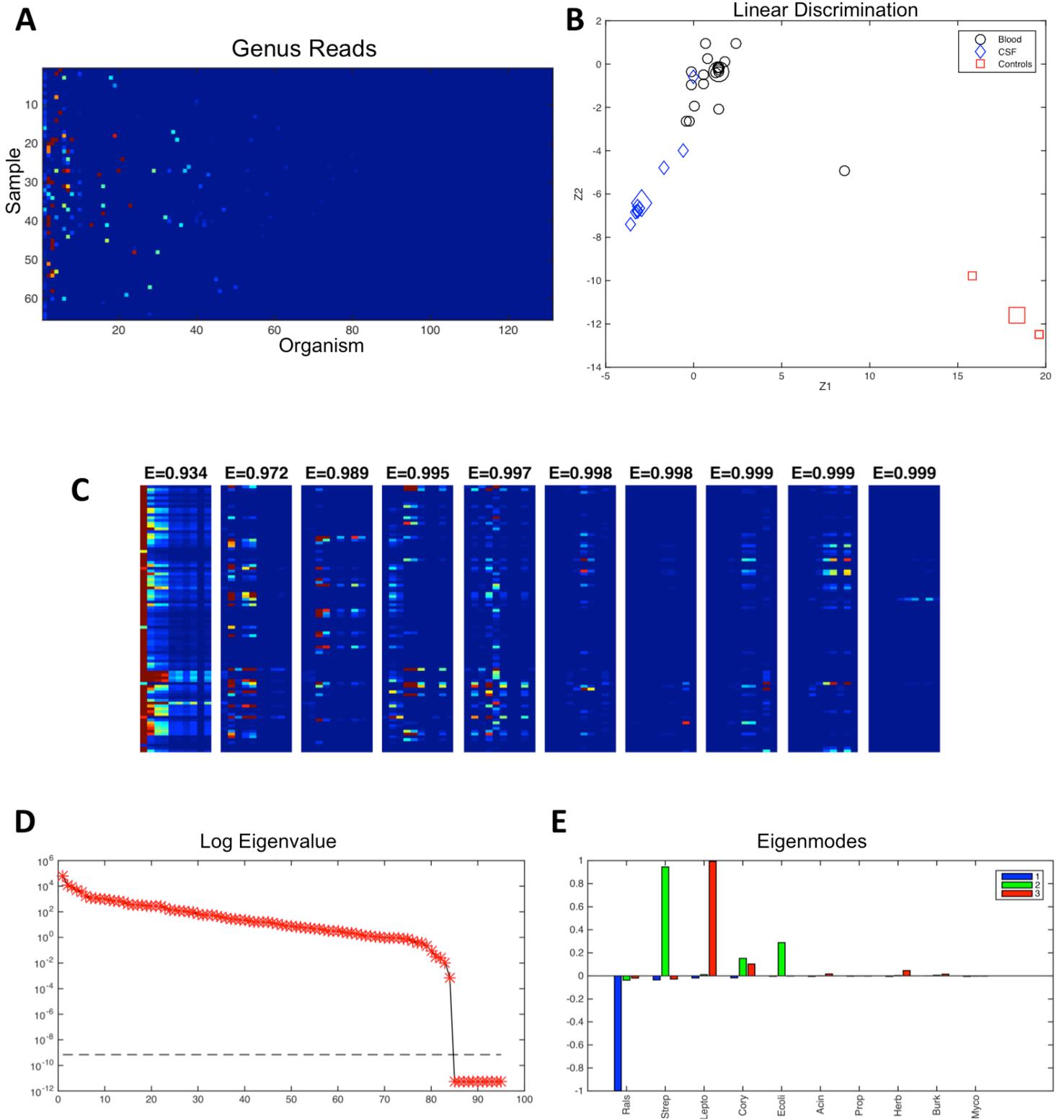

# Figure 2

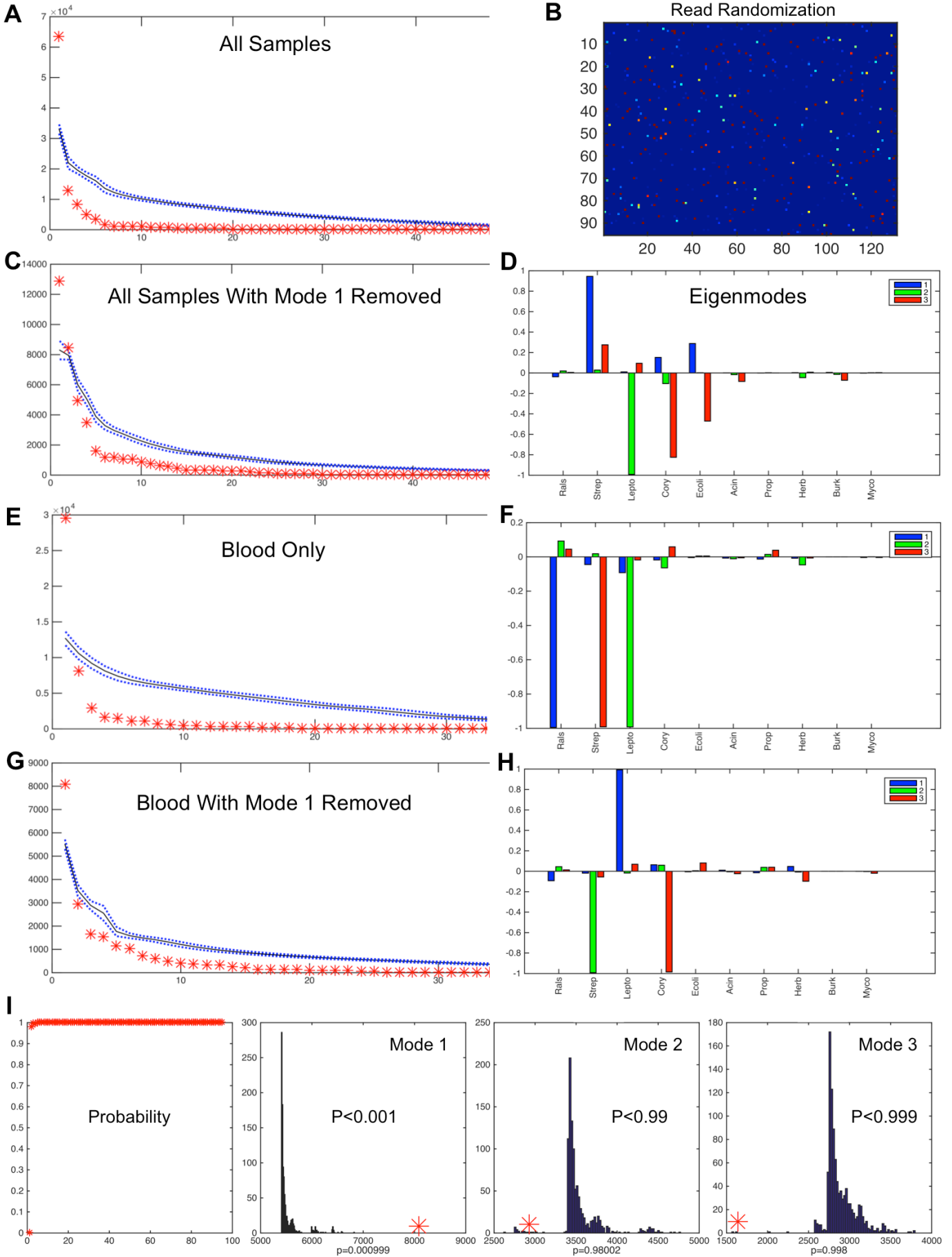